\documentclass[conference]{IEEEtran}
\IEEEoverridecommandlockouts
\usepackage{cite}
\usepackage{amsmath,amssymb,amsfonts}
\usepackage{algorithmic}
\usepackage{graphicx}
\usepackage{booktabs}
\usepackage{multirow}
\usepackage{makecell}
\usepackage{textcomp}
\usepackage{xcolor}
\def\BibTeX{{\rm B\kern-.05em{\sc i\kern-.025em b}\kern-.08em
    T\kern-.1667em\lower.7ex\hbox{E}\kern-.125emX}}
\begin{document}
\title{Task-Aware Bitrate Thresholds for Codec-Robust Ultrasound Segmentation
}

\author{
\IEEEauthorblockN{Alireza Siyavashi, Arman Shirzad, Christian Herglotz}
\IEEEauthorblockA{Chair of Computer Engineering, Brandenburgische Technische Universit\"at (BTU), Cottbus, Germany\\
Email: \{alireza.siyavashi, arman.shirzad, christian.herglotz\}@b-tu.de}
}
\IEEEaftertitletext{\vspace{-20pt}}
\maketitle
\begin{abstract}
Remote robot surgery and tele-ultrasound increasingly rely on real-time streaming under strict bandwidth constraints, making lossy compression unavoidable. 
Since the receiving end of these clinical pipelines is increasingly a machine rather than a human observer, codec selection must be guided by downstream task accuracy rather than perceptual fidelity.
We evaluate three widely deployed codecs H.264/AVC, HEVC, and AV1, on prostate ultrasound segmentation, reporting task-aware compression performance with region overlap (Dice) and boundary accuracy (HD95) as the quality metrics. 
To compare codecs fairly, we match their operating points at a common perceptual-quality (SSIM) level.
We establish, per codec, the lowest bitrate that preserves acceptable segmentation accuracy, reporting the 5th-percentile Dice alongside the mean to expose worst-case degradation.
All three codecs preserve segmentation quality (mean Dice $\geq$ 0.84) 
down to approximately 0.07–0.20 bits per voxel (bpv). 
Below this point, AV1's accuracy declines gradually, while H.264's drops sharply. 
As a result, H.264 needs 82\% more bitrate than AV1 to reach the same region-overlap accuracy (Dice), and 104.7\% more to reach the same boundary accuracy (HD95).
\end{abstract}

\begin{IEEEkeywords}
Telemedicine, Video streaming, Video compression, Ultrasound Imaging, Medical Image segmentation
\end{IEEEkeywords}
\begin{figure*}[t]
\centering
\includegraphics[width=\textwidth]{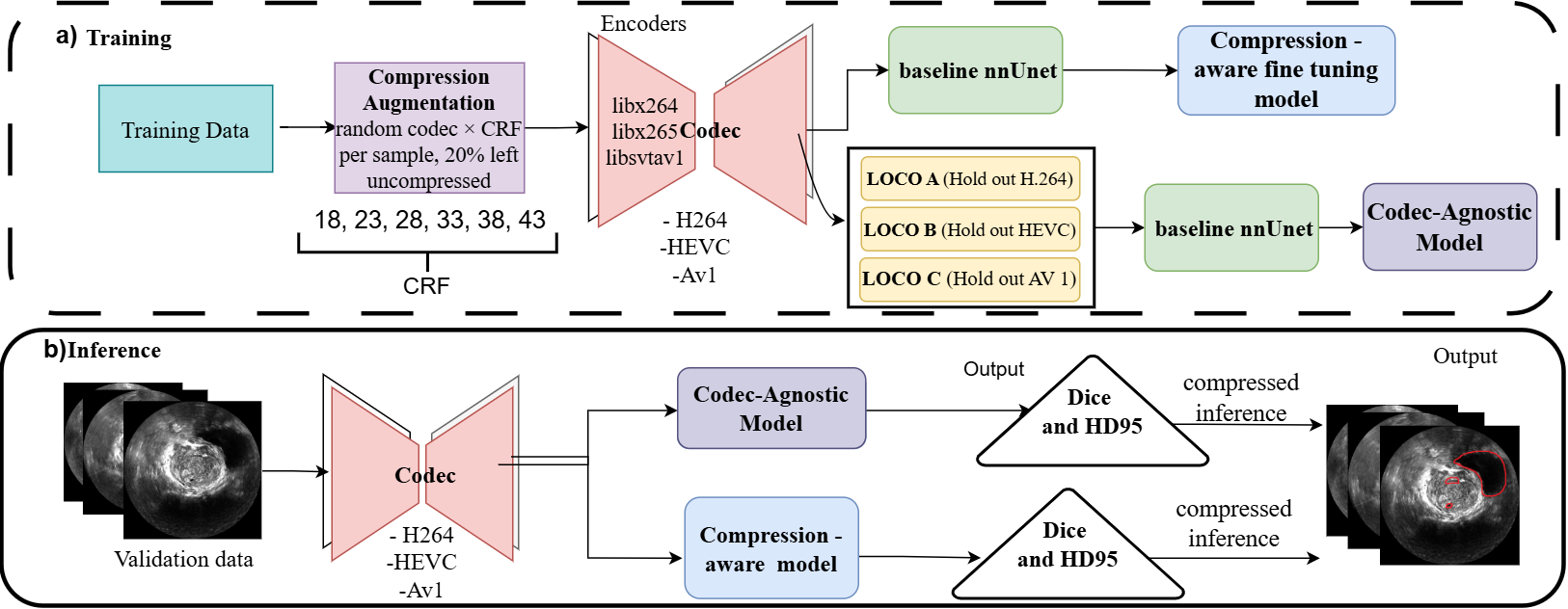}
\vspace{-6mm}
\caption{Overview of the proposed pipeline. (a)~Training: the baseline
nnU-Net is fine-tuned with codec augmentation across three
codecs and six CRF levels (20\% left uncompressed). The top path produces
the compression-aware model; the bottom path applies the
leave-one-codec-out (LOCO) protocol to produce three codec-agnostic
models. (b)~Inference: validation volumes are compressed and evaluated
using both models, reporting Dice and HD95.}
\label{fig:flowchart}
\vspace{-5mm}
\end{figure*}
\section{Introduction}
\label{Sec:introduction}
In recent years, remote surgery for medical practitioners has expanded rapidly, with robotic surgical platforms and tele-ultrasound systems enabling specialists to guide procedures across geographic boundaries~\cite{ding2026telerobotic,ebihara2022tele}. 
These workflows depend on real-time streaming of high-resolution medical imaging data, such as 3D ultrasound, under strict bandwidth constraints, making lossy compression unavoidable~\cite{panayides2020battle}.
As hospitals adopt AI-assisted pipelines where the receiving end is increasingly a machine, segmentation models driving surgical navigation or automated measurement of anatomical structures, the critical question shifts from how compression affects visual quality to how compression affects the neural network that acts on the decoded images~\cite{duan2020video,choi2022scalable}.
In Video Coding for Machines (VCM), the neural network's task accuracy is a more meaningful optimization target than classical quality metrics~\cite {duan2020video,choi2018deep}.

Several studies adopt this view for general-purpose vision: Pawlowski et al.\cite{pawlowski2025impact} report detection degradation under lossy coding using YOLOv7~\cite{wang2022efficient}.
Janeiro et al.\cite{janeiro2023visual} demonstrate that fine-tuning on compressed data recovers much of the lost recognition accuracy.
However, these works evaluate on 2D natural-image benchmarks using detection-oriented metrics, like Intersection over Union (IoU) and mean Average Precision (mAP),  and do not address a requirement critical to medical segmentation. 
In surgical navigation, a mask with inaccurate boundaries can misguide instrument placement; therefore, evaluation must capture both region overlap, Dice Similarity Coefficient score, and boundary accuracy 95th Percentile Hausdorff Distance (HD95)\cite{huttenlocher2002comparing}.

Existing studies are further limited in two distinct ways.
First, most consider only one or two codecs, often with a single encoder implementation, so their conclusions may not be generalized to other codecs or to other encoders of the same standard, offering limited guidance for hospitals selecting a codec configuration.
Second, they report only the mean accuracy across cases, which can mask substantially larger degradation in individual patients; for a safety-critical task, a worst-case or percentile-based criterion is more appropriate~\cite{rodrigues2025objective}.
While recent work has explored modifying the encoder's internal rate-distortion optimization to preserve neural-network features~\cite{fischer2020video, ge2024task}, these approaches require custom encoder builds that are unavailable in current clinical streaming infrastructure, where hospitals deploy off-the-shelf encoders with default settings.

In this paper, we present a three-codec (H.264, HEVC, AV1) task-aware rate-distortion study for ultrasound segmentation with nnU-Net~\cite{isensee2024nnu}, reporting Bj{\o}ntegaard-Delta rates (BD-Rate) with Dice and HD95 as quality axes.
We establish per-codec safe bitrate thresholds and report the 5th-percentile Dice to expose worst-case, patient-level degradation that the mean masks.
We further fine-tune nnU-Net on compressed data to improve low-bitrate robustness, and use a leave-one-codec-out (LOCO) protocol, fine-tuning on two codecs and testing on the held-out third, to assess generalization to unseen codecs.
To the best of our knowledge, this is the first cross-codec 
task-aware study for medical image segmentation. Our main 
contributions are:
\begin{itemize}
    \item Task-aware BD-Rate analysis across three codecs (H.264,
    HEVC, AV1), showing that H.264 needs 82\% more bitrate than AV1 at matched Dice.
    \item Per-codec safe bitrate thresholds, reporting worst-case (5th-percentile) Dice alongside the mean, as a deployable criterion for the lowest acceptable bitrate.
    \item A leave-one-codec-out (LOCO) protocol showing that robustness learned from two codecs generalizes to an unseen third, without codec-specific retraining.     
\end{itemize}
The remainder of this paper is organized as follows. Section~\ref{sec:background} reviews the codecs and segmentation framework. Section~\ref{sec:methodology} describes the experimental pipeline. Section~\ref{sec:evaluation} defines the evaluation metrics and protocols. Section~\ref{sec:result} presents and discusses the results. Section~\ref{sec:conclusion} concludes with deployment recommendations and future work.
\section{Background}
\label{sec:background}
We evaluate three codecs spanning successive coding generations: \textbf{H.264}~\cite{wiegand2003overview}, the most widely deployed clinical standard; \textbf{HEVC}~\cite{sullivan2012overview}, which roughly halves H.264's bitrate at equal quality and is increasingly used in medical video~\cite{panayides2020battle}; and \textbf{AV1}~\cite{chen2020av1}, a royalty-free Alliance for Open Media (AOM) codec that further improves low-to-medium-bitrate efficiency, encoded with libsvtav1~\cite{kossentini2020svtav1}. All encoders run under FFmpeg constant-quality (CRF) control.
We use nnU-Net~\cite{isensee2024nnu} as the backbone; its self-configuring pipeline minimizes manual tuning, so observed changes reflect compression rather than model configuration.
\vspace{-3mm}
\section{Methodology}
\label{sec:methodology}
\label{sec:pipeline}
Fig.~\ref{fig:flowchart} summarizes the pipeline: we (i)~train a baseline nnU-Net on uncompressed 3D Transrectal Ultrasound (TRUS) data; (ii)~compress the validation set with H.264, HEVC, and AV1 at six CRF levels (shown in Fig.~\ref{fig:flowchart}a); (iii)~evaluate the baseline and build rate-task curves on Dice and HD95; and (iv)~fine-tune nnU-Net with codec augmentation and a leave-one-codec-out (LOCO) protocol.

We use the public Prostate Magnetic Resonance Imaging (MRI)--Ultrasound (US) Biopsy dataset from The Cancer Imaging Archive (TCIA)~\cite{prostatetciav2}: 1761 3D Transrectal Ultrasound (TRUS) sweeps from 1151 patients.
We use only the grayscale 8-bit TRUS NIfTI volumes (Neuroimaging Informatics Technology Initiative) and prostate masks rasterized from the provided Standard Triangle Language (STL) meshes; MRI and biopsy data are not used.
Volume dimensions vary across cases (a representative volume is $436\times436\times281$ at 0.2848\,mm isotropic spacing); nnU-Net resamples all volumes to a common spacing during preprocessing.
\subsection{Compression}
\label{sec:compression}
To simulate bandwidth-limited streaming, each 3D volume is encoded as a grayscale video at a nominal 30\,fps, with the $Z$-axis as time; strongly correlated adjacent TRUS slices give inter-frame prediction genuine redundancy to exploit. 
Intensities fill the Y plane, while the U and V planes are set to the neutral YUV\,420 value (negligible bitrate) and discarded after decoding; odd dimensions are padded by border replication.
Since the frame rate only scales the bitrate in Mbps, we also report frame-rate-independent bits per voxel (bpv). Real streaming involves probe motion and tissue deformation, so the reported bitrates are lower bounds.

We compress the validation set with FFmpeg (version 8.0.1)~\cite{team2015complete} using \textbf{H.264} (libx264), \textbf{HEVC} (libx265), and \textbf{AV1} (libsvtav1), each at six Constant Rate Factor (CRF) settings $\mathcal{C} = \{18, 23, 28, 33, 38, 43\}$, yielding 18 configurations. 
Encoder parameters are fixed and not tuned per codec: libx264 and libx265 use the default \texttt{medium} preset with a 300-frame Group of Pictures (GOP) and scene-cut detection disabled, whereas libsvtav1 uses \texttt{preset 8} with its default keyframe interval, which is not matched to the 300-frame GOP.
A longer interval would slightly favour AV1 but is far too small an effect to explain the BD-Rate gap, and preset differences are normalized out by the SSIM harmonization below. Since CRF controls distortion rather than bitrate, we report the mean achieved bitrate over the 299 validation volumes. For a fair comparison at matched perceptual quality, we anchor on H.264 SSIM at CRF $\{18, 23, 28\}$ and find the AV1 CRFs with equivalent SSIM by linear interpolation, adding five AV1 operating points (CRF $\in \{30, 35, 36, 37, 41\}$). HEVC tracks H.264 SSIM within $\pm$0.5\,CRF at every anchor and needs no additional encodes.
\subsection{Baseline Segmentation}
\label{ssec:baseline}
We use the nnU-Net \texttt{3d\_fullres} configuration, trained on $N_{\text{train}} = 1462$ TRUS volumes and evaluated on a held-out set of $299$ volumes, split at the patient level so no patient appears in both. 
We use a single train/validation split rather than cross-validation, since we seek the compression threshold at which accuracy degrades significantly, not the best-performing model.
\vspace{-1.5mm}
\subsection{Compression-Aware Fine-Tuning}
\label{ssec:finetuning}
To improve robustness under compression, we fine-tune the baseline checkpoint with codec augmentation.
During training, each sample is either left uncompressed with probability $p_{\text{clean}}=0.20$ (to prevent catastrophic forgetting), or encoded and decoded using a randomly sampled codec-CRF pair drawn uniformly from the three codecs and the six-point CRF grid. 

All models were trained using stochastic gradient descent (SGD) with momentum($\mu$ = 0.99)~\cite{sutskever2013importance} and an initial learning rate of 0.01, decayed according to a polynomial schedule. Each epoch comprised 250 iterations. 
The compression-aware model was trained for 75 epochs and each leave-one-codec-out (LOCO) model for 100 epochs, with all models initialized from the baseline checkpoint.
Instead of training on a precomputed compressed corpus, compression was applied online to each sample during training.
We refer to this all-codec fine-tuned model as the compression-aware model.
To test whether compression robustness transfers across codecs, we additionally fine-tune three leave-one-codec-out (LOCO) models with the same augmentation, each holding out one codec from the compression pool:
\textbf{LOCO-A} trains on \{HEVC, AV1\} and holds out H.264;
\textbf{LOCO-B} trains on \{H.264, AV1\} and holds out HEVC;
\textbf{LOCO-C} trains on \{H.264, HEVC\} and holds out AV1.
Each LOCO model is fine-tuned from the baseline checkpoint, with the held-out codec never used to compress its training samples. At evaluation, each model is tested across all three codecs and six CRF levels: its accuracy on the held-out codec measures generalization to a codec unseen during fine-tuning, while its accuracy on the two training codecs confirms that this generalization does not degrade performance on the codecs it was trained on.
\begin{table*}[t]
\centering
\caption{Per-codec baseline rate-task performance across all six CRF settings. For each codec we report achieved bitrate (Mbps and bits per voxel, bpv), mean and 5th-percentile Dice, HD95, PSNR, and SSIM. Bold marks the best codec per CRF. For reference, on uncompressed data the model achieves a mean Dice of 0.869, a 5th-percentile Dice of 0.787, and an HD95 of 4.53 mm.}
\vspace{-2mm}
\label{tab:matched}
\footnotesize
\setlength{\tabcolsep}{2.5pt}
\begin{tabular}{@{}r ccccccc ccccccc ccccccc@{}}
\toprule
 & \multicolumn{7}{c}{H.264}
 & \multicolumn{7}{c}{HEVC}
 & \multicolumn{7}{c}{AV1}\\
\cmidrule(lr){2-8}\cmidrule(lr){9-15}\cmidrule(lr){16-22}
CRF
 & Mbps & bpv & Dice & 5\% & HD95 & PSNR & SSIM
 & Mbps & bpv & Dice & 5\% & HD95 & PSNR & SSIM
 & Mbps & bpv & Dice & 5\% & HD95 & PSNR & SSIM\\
\midrule
18 & 2.020 & 0.418 & 0.864 & 0.787 & \textbf{4.67} & 39.8 & 0.957
   & 1.872 & 0.391 & 0.857 & 0.770 & 4.93 & 40.0 & 0.955
   & 2.055 & 0.431 & \textbf{0.865} & \textbf{0.790} & 4.76 & 40.9 & 0.957\\
23 & 0.850 & 0.175 & 0.860 & 0.781 & 5.05 & 36.4 & 0.918
   & 0.858 & 0.179 & 0.854 & 0.763 & 5.22 & 37.0 & 0.920
   & 1.485 & 0.312 & \textbf{0.864} & \textbf{0.786} & \textbf{4.70} & 39.6 & 0.946\\
28 & 0.347 & 0.071 & 0.844 & 0.761 & 6.13 & 33.7 & 0.866
   & 0.365 & 0.076 & 0.842 & 0.753 & 5.62 & 34.6 & 0.870
   & 0.962 & 0.202 & \textbf{0.862} & \textbf{0.787} & \textbf{4.85} & 38.0 & 0.928\\
33 & 0.149 & 0.031 & 0.801 & 0.687 & 8.22 & 31.4 & 0.804
   & 0.155 & 0.032 & 0.811 & 0.711 & 7.18 & 32.4 & 0.811
   & 0.604 & 0.127 & \textbf{0.850} & \textbf{0.768} & \textbf{5.41} & 36.6 & 0.907\\
38 & 0.069 & 0.014 & 0.745 & 0.595 & 10.24 & 29.3 & 0.727
   & 0.069 & 0.014 & 0.771 & 0.654 & 9.08 & 30.4 & 0.733
   & 0.373 & 0.078 & \textbf{0.839} & \textbf{0.758} & \textbf{5.60} & 35.3 & 0.883\\
43 & 0.035 & 0.007 & 0.626 & 0.383 & 12.68 & 26.9 & 0.577
   & 0.036 & 0.007 & 0.697 & 0.518 & 14.12 & 28.3 & 0.643
   & 0.235 & 0.049 & \textbf{0.838} & \textbf{0.759} & \textbf{6.06} & 34.2 & 0.857\\
\bottomrule
\end{tabular}
\vspace{-5mm}
\end{table*}
\section{Evaluation Metrics}
\label{sec:evaluation}
\begin{figure*}[t]
\centering
\includegraphics[width=\textwidth]{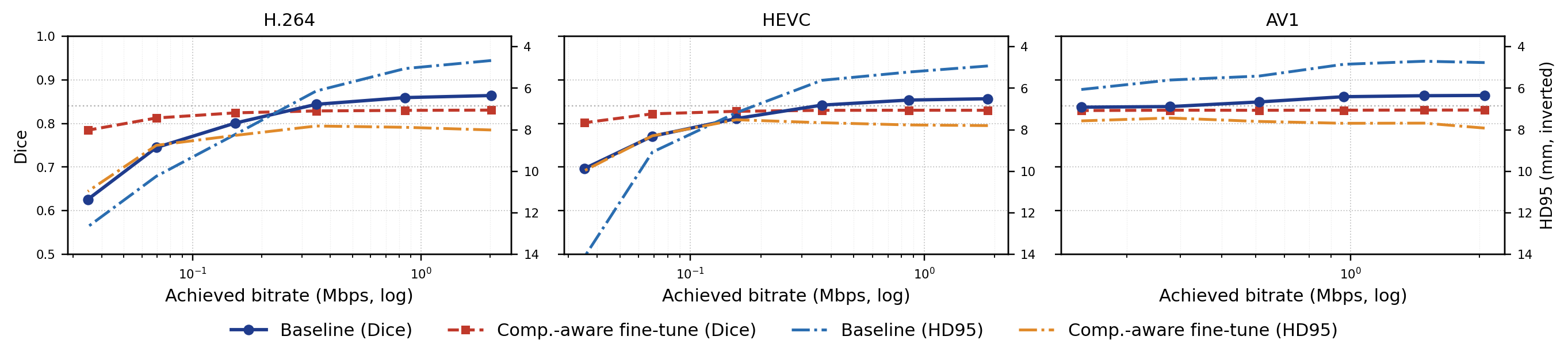}
\vspace{-8mm}
\caption{Rate-task curves per codec. Solid lines: Dice (left axis); 
dash-dot lines: HD95 (right axis, inverted, lower is better plotted 
upward). Blue: baseline; red: compression-aware fine-tuned.}
\label{fig:ratetask}
\vspace{-6mm}
\end{figure*}
Segmentation quality is measured using two complementary metrics.
The Dice Similarity Coefficient (DSC)~\cite{reinke2024pitfalls} quantifies volumetric overlap between predicted and ground-truth 
masks, ranging from 0 (no overlap) to 1 (perfect agreement); it is monotonically related to IoU via $\text{DSC} = 2\,\text{IoU}/(1+\text{IoU})$.
The 95th-percentile Hausdorff Distance (HD95)~\cite{huttenlocher2002comparing} measures boundary accuracy by computing the 95th percentile of the symmetric surface distances 
in mm between predicted and ground-truth contours, making it 
sensitive to boundary errors that Dice alone cannot capture~\cite{podobnik2024hdilemma}.
For completeness, we additionally report PSNR and SSIM between the original and decoded volumes.

Cross-codec efficiency is quantified using the Bjøntegaard-Delta Rate (BD-Rate)~\cite{herglotz2024bible}, computed over the overlapping quality range of the (log-bitrate, task-metric) curves.
Negative BD-Rate indicates that the test codec achieves the same task accuracy at lower bitrate.
We compute BD-Rate for both Dice and HD95 quality axes across all pairwise codec comparisons.
Dice is fit and interpolated directly, while HD95 is negated ($-\text{HD95}$) before fitting so the standard increasing-metric formulation applies according to Bj{\o}ntegaard formulation.

To provide guidance for practical applications, we define per-codec \emph{safe bitrate thresholds} as the minimum bitrate at which the mean Dice across all 299 test volumes remains above an acceptable floor (Dice $\geq 0.84$), i.e.\ no more than 3 percentage points below the uncompressed baseline testset(Dice\,=\,0.869).
This margin is consistent with inter-observer variability reported for prostate gland delineation in ultrasound~\cite{lenfant2025defining}.
To expose worst-case, patient-level behavior that the mean would otherwise mask, we additionally report the 5th-percentile Dice at every operating point.
\begin{figure}[t]
\centering
\includegraphics[width=\columnwidth]{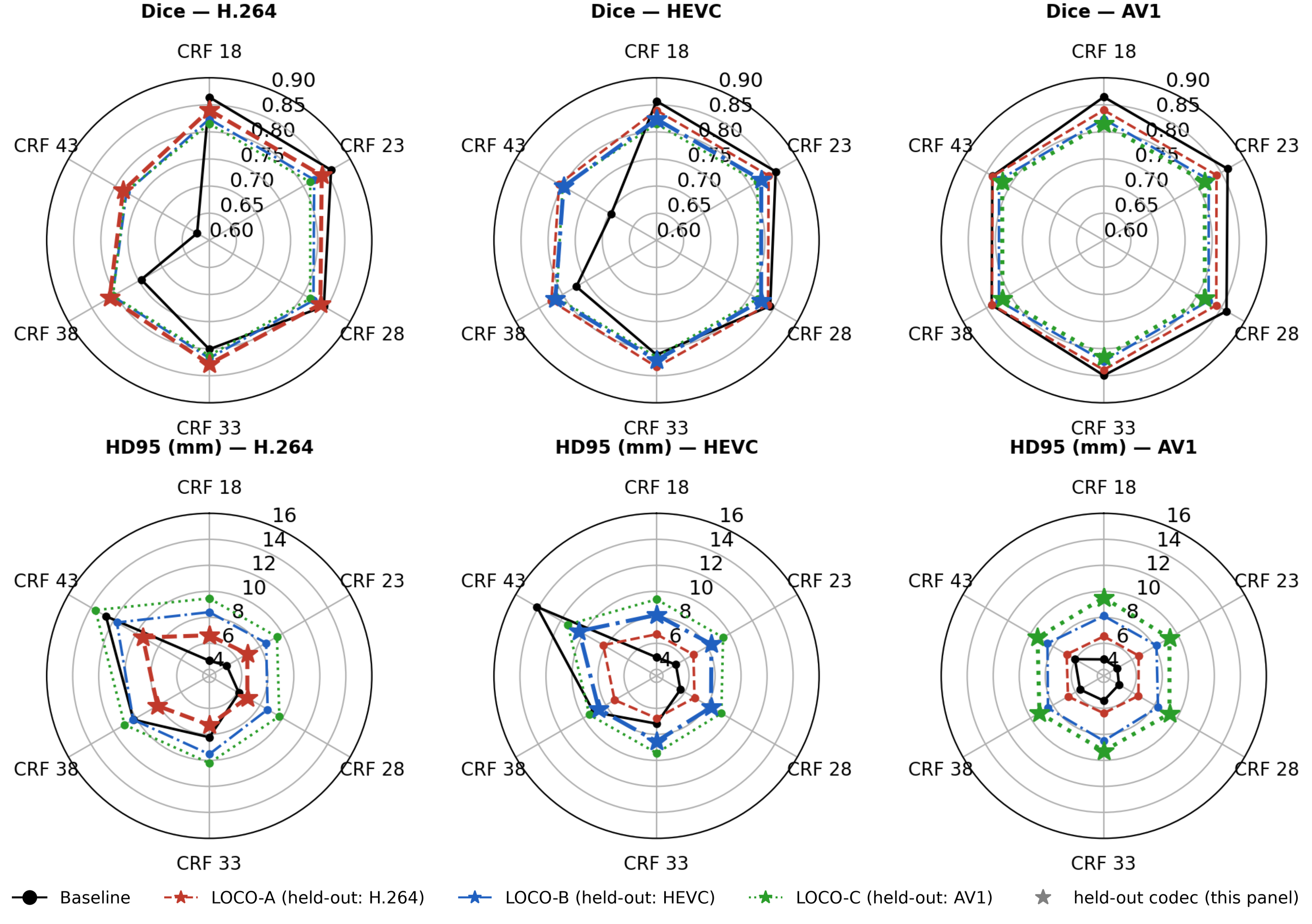}
\caption{Leave-one-codec-out (LOCO) generalization across the six-CRF
grid (top: Dice; bottom: HD95 in mm; one codec per column). Every LOCO
model stays near-flat on its held-out codec (stars), while the baseline
collapses at H.264 CRF\,43.}
\vspace{-6mm}
\label{fig:radar}
\end{figure}
\section{Results}
\label{sec:result}
\subsection{Baseline Rate-Task Performance}
Table~\ref{tab:matched} and Fig.~\ref{fig:ratetask} summarize baseline
segmentation accuracy as a function of bitrate across the three
codecs. 
At high bitrates ($\geq$\,1\,Mbps), all three codecs perform
comparably, with mean Dice between 0.857 and 0.865 and HD95 between
4.67 and 4.93\,mm.
Differences emerge as the bitrate decreases.
At a comparably low bitrate ($\sim$0.2\,Mbps), AV1 (CRF\,43, 0.235\,Mbps) retains a Dice of 0.838 and HD95 of 6.06\,mm, while
H.264 (CRF\,33, 0.149\,Mbps) drops to 0.801 and 8.22\,mm, a gap of 3.7
percentage points in Dice and 2.16\,mm in HD95.
At these operating points, the 5th-percentile Dice reveals a sharper contrast: AV1
maintains 0.759 versus H.264's 0.687, indicating that the worst-case
patient experiences over 7 percentage points better segmentation with
AV1. 
Task-aware BD-Rate analysis confirms this trend: H.264 requires 82.2\% more bitrate than AV1 to achieve the same Dice, and 104.7\% more for matched HD95. 
HEVC lies between H.264 and AV1, needing 54.1\% (Dice) and 52.5\% (HD95) more bitrate than AV1.

Holding mean Dice $\geq 0.84$, all three codecs reach a similar bitrate floor
(H.264 0.35\,Mbps, 0.071\,bpv, at CRF\,28, HEVC 0.37\,Mbps,  0.076\,bpv, at CRF\,28, AV1
0.42\,Mbps, 0.127\,bpv, at CRF\,37 on the SSIM-harmonized grid, refined from a
coarse-grid estimate of 0.61\,Mbps, 0.202\,bpv). 
The decisive difference appears
below this bitrate floor: AV1 degrades slowly, holding Dice $\approx$\,0.84
down to 0.235\,Mbps, 0.049\,bpv, (CRF\,43), whereas H.264 has already dropped to 0.80 at 0.15\,Mbps, 0.031\,bpv and continues to fall steeply.
Consistent with the
BD-Rate result, where H.264 needs 82\% more bitrate than AV1 for
matched Dice, AV1's advantage is graceful low-bitrate behavior rather than a lower minimum bitrate: under variable or constrained bandwidth, it maintains usable segmentation where H.264 fails, making it the preferred codec for such deployments.

Compression-aware fine-tuning (Fig.~\ref{fig:ratetask}, red curves) substantially mitigates the H.264 low-bitrate cliff, shifting Dice at the lowest operating point from approximately 0.63 to 0.79 and yielding a more uniform HD95 profile across codecs. Because the fine-tuned curves are nearly flat ($\approx$\,0.83 at every bitrate), this gain comes at a small cost at high bitrate, trading peak accuracy for low-bitrate robustness.

\subsection{LOCO Generalization}
Fig.~\ref{fig:radar} reports the LOCO results as radar plots, one panel per codec, with the six spokes corresponding to the CRF levels $\{18, 23, 28, 33, 38, 43\}$.
Each colored polygon is one model (the baseline and the three LOCO models), and a star marks the codec held out from that LOCO model's fine-tuning. 
The two rows are read in opposite directions: in the top row (Dice), a larger polygon, farther from the center, indicates better region overlap; in the bottom row (HD95), a smaller polygon, closer to the center, indicates better boundary accuracy, since lower HD95 is better.

We can observe that averaged over the six CRF levels, each LOCO model generalizes to its held-out codec, staying within a few Dice points of the baseline: LOCO-A reaches 0.823 on H.264 (vs.\ 0.790 baseline) and LOCO-B 0.817 on HEVC (vs.\ 0.805), both improving on the baseline, while LOCO-C reaches 0.816 on AV1 (vs.\ 0.853), just below the already-robust AV1 baseline. The improvement is largest on H.264, whose low-bitrate cliff is poorly handled by the baseline.
The contrast is sharpest at the H.264 CRF\,43 point: the baseline collapses (Dice 0.626, HD95 12.7\,mm), whereas all three LOCO models recover to Dice $\approx$\,0.78, including LOCO-A, which never saw H.264 during fine-tuning.
Because each LOCO model generalizes to a codec unseen during fine-tuning, this transfer can eliminate the need to retrain for every deployment codec.
\vspace{-3mm}
\section{Conclusion}
\label{sec:conclusion}
We presented a task-aware rate-distortion study of prostate ultrasound
segmentation across three codec families (H.264, HEVC, AV1) using
Dice and HD95 as quality axes. Our results show that AV1 is
substantially more bitrate-efficient than H.264 and HEVC for
preserving segmentation accuracy, and that per-codec safe bitrate thresholds, with worst-case 5th-percentile Dice reported alongside the mean, give concrete deployment guidance.
Compression-aware fine-tuning recovers
segmentation accuracy at extreme compression, and the
leave-one-codec-out protocol confirms that this robustness
generalizes to unseen codecs without retraining.

Future work will extend this framework in two directions. First, we
plan to evaluate VVC and optimize its configuration for medical
segmentation tasks. Second, since compression sensitivity is likely
modality-dependent, we aim to validate the proposed methodology on
other imaging modalities such as CT and MRI, and across different
anatomical targets.
\bibliographystyle{IEEEtran}
\bibliography{refrence}

@article{ding2026telerobotic,
  title     = {Telerobotic surgery: a comprehensive two-decade evolution and the integration of emerging technologies},
  author    = {Ding, Y. and Wang, S. and Lan, R. and Lin, W. and Liu, X. and He, W.},
  journal   = {International Journal of Surgery},
  volume    = {112},
  number    = {1},
  pages     = {1652--1672},
  year      = {2026},
  publisher = {LWW}
}

@article{choi2022scalable,
  title={Scalable image coding for humans and machines},
  author={Choi, Hyomin and Baji{\'c}, Ivan V},
  journal={IEEE Transactions on Image Processing},
  volume={31},
  pages={2739--2754},
  year={2022},
  publisher={IEEE}
}

@article{duan2020video,
  title={Video coding for machines: A paradigm of collaborative compression and intelligent analytics},
  author={Duan, Lingyu and Liu, Jiaying and Yang, Wenhan and Huang, Tiejun and Gao, Wen},
  journal={IEEE Transactions on Image Processing},
  volume={29},
  pages={8680--8695},
  year={2020},
  publisher={IEEE}
}

@inproceedings{pawlowski2025impact,
  title={Impact of lossy compression on the precision of image analysis using deep neural networks},
  author={Paw{\l}owski, Pawe{\l} and Lorkiewicz, Jan and Balcerek, Julian},
  booktitle={2025 Signal Processing: Algorithms, Architectures, Arrangements, and Applications (SPA)},
  pages={82--87},
  year={2025},
  organization={IEEE}
}

@article{janeiro2023visual,
  title={Are visual recognition models robust to image compression?},
  author={Janeiro, Jo{\~a}o Maria and Frolov, Stanislav and El-Nouby, Alaaeldin and Verbeek, Jakob},
  journal={arXiv preprint arXiv:2304.04518},
  year={2023}
}

@inproceedings{choi2018deep,
  title={Deep feature compression for collaborative object detection},
  author={Choi, Hyomin and Baji{\'c}, Ivan V},
  booktitle={2018 25th IEEE International Conference on Image Processing (ICIP)},
  pages={3743--3747},
  year={2018},
  organization={IEEE}
}

@article{huttenlocher2002comparing,
  title={Comparing images using the Hausdorff distance},
  author={Huttenlocher, Daniel P and Klanderman, Gregory A. and Rucklidge, William J},
  journal={IEEE Transactions on pattern analysis and machine intelligence},
  volume={15},
  number={9},
  pages={850--863},
  year={2002},
  publisher={IEEE}
}

@article{rodrigues2025objective,
  title={Objective quality assessment of medical images and videos: review and challenges},
  author={Rodrigues, Rafael and L{\'e}v{\^e}que, Lucie and Guti{\'e}rrez, Jes{\'u}s and Jebbari, Houda and Outtas, Meriem and Zhang, Lu and Chetouani, Aladine and Al-Juboori, Shaymaa and Martini, Maria G and Pinheiro, Antonio MG},
  journal={Multimedia Tools and Applications},
  volume={84},
  number={25},
  pages={29915--29948},
  year={2025},
  publisher={Springer}
}

@article{ebihara2022tele,
  title={Tele-assessment of bandwidth limitation for remote robotics surgery},
  author={Ebihara, Yuma and Oki, Eiji and Hirano, Satoshi and Takano, Hironobu and Ota, Mitsuhiko and Morohashi, Hajime and Hakamada, Kenichi and Urushidani, Shigeo and Mori, Masaki},
  journal={Surgery today},
  volume={52},
  number={11},
  pages={1653--1659},
  year={2022},
  publisher={Springer}
}

@article{panayides2020battle,
  title={The battle of the video codecs in the healthcare domain-a comparative performance evaluation study leveraging VVC and AV1},
  author={Panayides, Andreas S and Pattichis, Marios S and Pantziaris, Marios and Constantinides, Anthony G and Pattichis, Constantinos S},
  journal={IEEE Access},
  volume={8},
  pages={11469--11481},
  year={2020},
  publisher={IEEE}
}

@inproceedings{ge2024task,
  title={Task-aware encoder control for deep video compression},
  author={Ge, Xingtong and Luo, Jixiang and Zhang, Xinjie and Xu, Tongda and Lu, Guo and He, Dailan and Geng, Jing and Wang, Yan and Zhang, Jun and Qin, Hongwei},
  booktitle={Proceedings of the IEEE/CVF Conference on Computer Vision and Pattern Recognition},
  pages={26036--26045},
  year={2024}
}

@inproceedings{fischer2020video,
  title={Video coding for machines with feature-based rate-distortion optimization},
  author={Fischer, Kristian and Brand, Fabian and Herglotz, Christian and Kaup, Andr{\'e}},
  booktitle={2020 IEEE 22nd International Workshop on Multimedia Signal Processing (MMSP)},
  pages={1--6},
  year={2020},
  organization={IEEE}
}

@article{wang2022efficient,
  title     = {Efficient detection model of steel strip surface defects based on {YOLO-V7}},
  author    = {Wang, Y. and Wang, H. and Xin, Z.},
  journal   = {IEEE Access},
  volume    = {10},
  pages     = {133936--133944},
  year      = {2022},
  publisher = {IEEE}
}

@inproceedings{isensee2024nnu,
  title        = {{nnU-Net} revisited: A call for rigorous validation in 3D medical image segmentation},
  author       = {Isensee, F. and Wald, T. and Ulrich, C. and Baumgartner, M. and Roy, S. and Maier-Hein, K. and Jaeger, P. F.},
  booktitle    = {International Conference on Medical Image Computing and Computer-Assisted Intervention},
  pages        = {488--498},
  year         = {2024},
  organization = {Springer}
}

@article{chen2020av1,
  author  = {Chen, Yue and Mukherjee, Debargha and Han, Jingning 
             and Grange, Adrian and Xu, Yaowu and Parker, Sarah 
             and Chen, Cheng and Su, Hui and Joshi, Urvang and 
             Chiang, Ching-Han and others},
  title   = {An Overview of Coding Tools in {AV1}: The First Video 
             Codec from the {Alliance for Open Media}},
  journal = {APSIPA Transactions on Signal and Information Processing},
  volume  = {9},
  pages   = {e21},
  year    = {2020}
}

@inproceedings{kossentini2020svtav1,
  author    = {Kossentini, Faouzi and Guermazi, Hassen and 
               Mahdi, Nader and Nouira, Chekib and 
               Naghdinezhad, Amir and Tmar, Hassene and 
               Khlif, Omar and Worth, Phoenix and 
               Ben Amara, Foued},
  title     = {The {SVT-AV1} Encoder: Overview, Features and 
               Speed-Quality Tradeoffs},
  booktitle = {Proc. SPIE Applications of Digital Image 
               Processing XLIII},
  volume    = {11510},
  year      = {2020}
}

@misc{prostatetciav2,
  title        = {Prostate {MRI} and {Ultrasound} {With} {Pathology} and {Coordinates} of {Tracked} {Biopsy} ({Prostate-MRI-US-Biopsy}) (version 2)},
  author       = {Natarajan, S. and Priester, A. and Margolis, D. and Huang, J. and Marks, L.},
  year         = {2020},
  publisher    = {The Cancer Imaging Archive},
  howpublished = {Dataset},
  doi          = {10.7937/TCIA.2020.A61IOC1A}
}

@article{wiegand2003overview,
  author  = {Wiegand, Thomas and Sullivan, Gary J. and 
             Bjontegaard, Gisle and Luthra, Ajay},
  title   = {Overview of the {H.264/AVC} Video Coding Standard},
  journal = {IEEE Trans. Circuits and Systems for Video Technology},
  volume  = {13},
  number  = {7},
  pages   = {560--576},
  year    = {2003}
}

@article{sullivan2012overview,
  author  = {Sullivan, Gary J. and Ohm, Jens-Rainer and 
             Han, Woo-Jin and Wiegand, Thomas},
  title   = {Overview of the {High Efficiency Video Coding (HEVC)} 
             Standard},
  journal = {IEEE Trans. Circuits and Systems for Video Technology},
  volume  = {22},
  number  = {12},
  pages   = {1649--1668},
  year    = {2012}
}

@article{reinke2024pitfalls,
  title   = {Understanding metric-related pitfalls in image analysis validation},
  author  = {Reinke, A. and Tizabi, M. D. and Baumgartner, M. and S{\'a}nchez, C. I. and {Metrics Reloaded}},
  journal = {Nature Methods},
  volume  = {21},
  number  = {2},
  pages   = {182--194},
  year    = {2024},
  doi     = {10.1038/s41592-023-02150-0}
}

@inproceedings{podobnik2024hdilemma,
  title     = {{HDilemma}: Are Open-Source Hausdorff Distance Implementations Equivalent?},
  author    = {Podobnik, Ga{\v{s}}per and Vrtovec, Toma{\v{z}}},
  booktitle = {Medical Image Computing and Computer Assisted Intervention -- MICCAI 2024},
  series    = {Lecture Notes in Computer Science},
  volume    = {15009},
  pages     = {308--317},
  publisher = {Springer},
  year      = {2024},
  doi       = {10.1007/978-3-031-72114-4_30}
}

@article{herglotz2024bible,
  author  = {Herglotz, Christian and Och, Holger and Meyer, 
             Andreas and Ramasubbu, Georg and others},
  title   = {The {Bj{\o}ntegaard} Bible: Why Your Way of 
             Comparing Video Codecs May Be Wrong},
  journal = {IEEE Trans. Image Processing},
  volume  = {33},
  pages   = {987--1001},
  year    = {2024}
}

@misc{team2015complete,
  title={A Complete, Cross-Platform Solution to Record, Convert, and Stream Audio and Video},
  author={FFmpeg Team},
  year={2015}
}

@article{lenfant2025defining,
  title={Defining ground truth for prostate segmentation of transrectal ultrasound images: Inter-and intra-observer variability of manual versus semi-automatic methods},
  author={Lenfant, Louis and Beitone, Cl{\'e}ment and Troccaz, Jocelyne and Fiard, Gaelle and Malavaud, Bernard and Voros, Sandrine and Mozer, Pierre C},
  journal={Medical Physics},
  volume={52},
  number={8},
  pages={e18025},
  year={2025},
  publisher={Wiley Online Library}
}

@inproceedings{sutskever2013importance,
  author    = {Sutskever, Ilya and Martens, James and Dahl, George and Hinton, Geoffrey},
  title     = {On the importance of initialization and momentum in deep learning},
  booktitle = {Proceedings of the 30th International Conference on Machine Learning (ICML)},
  pages     = {1139--1147},
  year      = {2013}
}

\end{document}